\documentclass[11pt,twoside]{article}
\usepackage{asp2010}

\resetcounters

\begin{document}

\title{Astrophysics Source Code Library Enhancements}
\author{Robert J. Hanisch$^1$, Alice Allen$^2$, G. Bruce Berriman$^3$, Kimberly DuPrie$^{2,4}$, Jessica Mink$^5$, Robert J. Nemiroff$^6$, Judy Schmidt$^2$, Lior Shamir$^7$, Keith Shortridge$^8$, Mark Taylor$^9$,  Peter J. Teuben$^{10}$, and John Wallin$^{11}$
\affil{$^1$National Institute of Standards and Technology}
\affil{$^2$Astrophysics Source Code Library}
\affil{$^3$Infrared Processing and Analysis Center, California Institute of Technology}
\affil{$^4$Space Telescope Science Institute}
\affil{$^5$Smithsonian Astrophysical Observatory}
\affil{$^6$Michigan Technological University}
\affil{$^7$Lawrence Technological University}
\affil{$^8$Australian Astronomical Observatory}
\affil{$^9$University of Bristol}
\affil{$^{10}$University of Maryland}
\affil{$^{11}$Middle Tennessee State University}}

\begin{abstract}
The Astrophysics Source Code Library (ASCL)\footnote{\url{http://ascl.net}} is a free online registry of codes used in astronomy research; it currently contains over 900 codes and is indexed by ADS. The ASCL has recently moved a new infrastructure into production. The new site provides a true database for the code entries and integrates the WordPress news and information pages and the discussion forum into one site. Previous capabilities are retained and permalinks to ascl.net continue to work. This improvement offers more functionality and flexibility than the previous site, is easier to maintain, and offers new possibilities for collaboration. This presentation covers these recent changes to the ASCL.

\end{abstract}

\section{Introduction}
Started in 1999, the Astrophysics Source Code Library (ASCL) is a free online registry for source codes used in astronomy and astrophysics. It seeks to make scientist-written software used in research more discoverable for examination and thus improve the transparency of research. As of September 2014, the resource has over 900 entries and is indexed by the SAO/NASA Astrophysics Data System (ADS). 

\section{ASCL Versions 1 and 2}
ASCL entries were initially housed on straight HTML pages, one code per webpage; as the resource required code be deposited, any software in the ASCL could be downloaded from its entry webpage. Other more general HTML pages categorized the codes, explained the purpose of the resource, provided information on submitting software and links of possible interest to viewers, and listed changes to the site. This ASCL version 1 eventually stored 37 codes. An unsuccessful search for a new editor in 2003 brought growth of the library to a stop, though the existing entries continued to be served.  

In 2010, the resource moved to a phpbb forum on the Astronomy Picture of the Day (APOD) discussion site Starship Asterisk\footnote{\url{http://asterisk.apod.com/index.php}}; the ASCL forum was among other forums unrelated to research software. Also in 2010, a new editor was found and in 2011, an advisory committee was formed. The requirement for housing an archive file of registered software was dropped, and though the ASCL can and does house codes, most entries point to a download site elsewhere. As the code library grew, so did its technical needs. A WordPress site was developed and in February 2013 became the new front end. This provided a more polished entry point to the ASCL, but also created a cumbersome two-site system; the two sites were linked but not fully integrated with one another. 

\section{ASCL Version 3: User Benefits}
Early in 2014, work started on an infrastructure that would better serve the growing registry; in July 2014, ASCL version 3 was moved into production. The new site utilizes a MySQL database and was built using the open source PHP framework Codeigniter by EllisLab. WordPress, used for content management, is fully integrated into the site. The discussion forum was replicated from Starship Asterisk and is also integrated into the site. These changes pull all the disparate pieces the ASCL had been using together for a seamless user experience. For the user, the new site offers the following enhancements: \\ \\
{\em Newest codes on home page}\\
The titles and descriptions of the ten most recent additions to the ASCL are shown in reverse date order on the home page and link to their full entries. \\ \\
{\em Improved submission process}\\
Though the phpbb ASCL site allowed an author to submit a code, the process was not intuitive. The new infrastructure offers a form for submissions that provides more guidance to the submitter. New submissions appear immediately on the ASCL and trigger an entry for the software on the ASCL discussion forum. Authors and other interested parties can if they wish subscribe to the discussion thread for any updates to the thread that may occur. Though the discussion forum is not used much now, we hope it will grow into a resource for software authors and users alike. \\ \\
{\em More flexible browsing}\\
Users can browse the ASCL by date and alphabetically by code name, and the order for these browsing options can be reversed. Browsing offers two display modes, one that shows just ascl ID and code name or title, and a longer view that includes the abstract for the software. The number of entries displayed can also be changed. By default, the ASCL shows 100 codes per page when browsing; this can be changed to show 50, 250, or all entries on the page instead. \\ \\
{\em One-click author searching}\\
As with arXiv, author names are links and clicking on an author link brings up other entries, if any, for software written by that author. The resulting page also provides an opportunity to refine or broaden the search by, for example, searching only for a last name. \\ \\
{\em Associated papers}\\
ASCL version 3 expands the ability to list research that uses a code even if the paper does not formally cite it. As many older papers do not cite software, this can over time provide evidence of the impact a code has, thus helping to demonstrate its value. These associations are passed to ADS to improve linking between papers and research-enabling software. \\ \\
{\em RSS feed}\\ 
The ASCL now includes an RSS feed for updates to the resource.  \\ 

\section{ASCL Version 3: Editor Benefits}
The upgraded infrastructure benefits the ASCL editors, too, automating tasks that previously were done by hand and adding features to help maintain the registry. For editors, the site provides:\\ \\
{\em Automatic bibcode generation}\\
Previously, bibcodes were not stored on the ASCL; ADS created bibcodes for ASCL entries after the entries were ingested by that system. Now, the ASCL system generates an entry's bibcode and, after the entry has been verified and given an ascl ID by an editor, passes it to ADS with other information. \\ \\
{\em Automatic ascl ID assignment}\\
ASCL IDs had been assigned manually by an editor; these are now automatically assigned after an editor has approved a code entry. \\ \\
{\em On demand dynamic record formatting for ADS ingestion}\\
ADS and ASCL representatives met once and exchanged many emails while the new infrastructure was being developed, as one desired outcome of the development effort was smoother integration between ADS and the ASCL. Whereas previously entries to be uploaded to ADS were periodically formatted by hand in a flat text file and sent to ADS for uploading, ADS personnel can now pull new ASCL entries formatted to ADS specifications. This streamlines ADS ingestion of records and greatly reduces opportunities for erroneous input.\\ \\ 
{\em Automatic creation of code discussion thread}\\
Upon code publication, a related thread is created on the connected discussion forum, thus removing the need for an editor to create it. The discussion post created is updated automatically when changes to the information it displays are made in the main code entry. \\ \\
{\em Information tracking}\\
The new site stores information that had been tracked completely independent of the ASCL, such as corresponding authors' email addresses, making the information available to all editors for use and maintenance. \\ \\
{\em Better reports}\\
The SQL database provides a more flexible platform not only for maintaining and displaying data, but also for reporting, giving editors the opportunity to look at stored information in new ways. It is now easier to expose ASCL data, which provides opportunities for greater collaboration and mirroring of ASCL entries on other sites.

In addition to the above, ASCL version 3 addresses a complaint people had voiced about the phpbb-based site: the way it looked. Use of a phpbb discussion forum made the ASCL look unprofessional and unpolished; concern with its appearance was raised by journal editors and users alike. As Judy Schmidt, who developed our enhanced site, is also a designer, the ASCL's appearance has improved dramatically.

\section{Conclusion}

At last year's ADASS, we had expressed a desire for "a new infrastructure for code entries with additional features such as an API, better search capability, more professional look, and increased automation of the back end."\citep{2014ASPC..485..477A} We also said we wanted to retain some features, such as the ability for people to post comments and questions about a code and a service to notify those subscribed to a thread when an addition or change appears. We continue to work on the few missing items, but are very happy to have so many of our users' wants and ours satisfied with the first release of the new site.

\bibliography{P8-8}

\begin{thebibliography}{}
\expandafter\ifx\csname natexlab\endcsname\relax\def\natexlab#1{#1}\fi
\expandafter\ifx\csname url\endcsname\relax
  \def\url#1{\texttt{#1}}\fi
\expandafter\ifx\csname urlprefix\endcsname\relax\def\urlprefix{URL }\fi
\providecommand{\eprint}[2][]{\url{#2}}

\bibitem[{{Allen} et~al.(2014){Allen}, {Berriman}, {DuPrie}, {Hanisch}, {Mink},
  {Nemiroff}, {Shamir}, {Shortridge}, {Taylor}, {Teuben}, \&
  {Wallen}}]{2014ASPC..485..477A}
{Allen}, A., {Berriman}, B., {DuPrie}, K., {Hanisch}, R.~J., {Mink}, J.,
  {Nemiroff}, R.~J., {Shamir}, L., {Shortridge}, K., {Taylor}, M.~B., {Teuben},
  P., \& {Wallen}, J. 2014, in Astronomical Data Anaylsis Softward and Systems
  XXIII, edited by N.~{Manset}, \& P.~{Forshay}, vol. 485 of Astronomical
  Society of the Pacific Conference Series, 477

\end{thebibliography}

\end{document}